\documentclass[sigconf]{acmart}

\AtBeginDocument{%
  }

\copyrightyear{2026}
\acmYear{2026}
\setcopyright{cc}
\setcctype{by}
\acmConference[DIS Companion '26]{Designing Interactive Systems Conference}{June 13--17, 2026}{Singapore, Singapore}
\acmBooktitle{Designing Interactive Systems Conference (DIS Companion '26), June 13--17, 2026, Singapore, Singapore}
\acmDOI{10.1145/3802974.3809425}
\acmISBN{979-8-4007-2632-3/2026/06}

\usepackage{xcolor}
\newcommand{\add}[1]{\textcolor{black}{#1}} 

\newcommand{\q}[1]{\textit{``#1''}} 

\usepackage{makecell}

\begin{document}

\title{Exploring a Multimodal Chatbot as a Facilitator in Therapeutic Art Activity}
\author{Le Lin}
\affiliation{%
\department{Department of Computer Science}
  \institution{City University of Hong Kong}
    \city{Hong Kong SAR}
  \country{China}
}
\email{lelin3-c@my.cityu.edu.hk}

\author{Zihao Zhu}
\affiliation{%
  \department{Department of Computer Science}
  \institution{City University of Hong Kong}
    \city{Hong Kong SAR}
  \country{China}
}
\email{zihaozhu9-c@my.cityu.edu.hk}

\author{Rainbow Tin Hung Ho}
\affiliation{%
  \department{Centre on Behavioral Health}
  \institution{The University of Hong Kong}
     \city{Hong Kong SAR}
  \country{China}
}
\email{tinho@hku.hk}

\author{Jing Liao}
\authornote{Corresponding author.}
\affiliation{%
  \department{Department of Computer Science}
  \institution{City University of Hong Kong}
    \city{Hong Kong SAR}
  \country{China}
}
\email{jingliao@cityu.edu.hk}

\author{Yuhan Luo}
\authornotemark[1]
\affiliation{%
  \department{Department of Computer Science}
  \institution{City University of Hong Kong}
    \city{Hong Kong SAR}
  \country{China}
}
\email{yuhanluo@cityu.edu.hk}

\renewcommand{\shortauthors}{Lin et al.}

\begin{abstract}
 Therapeutic art activities, such as expressive drawing and painting, require the synergy between creative visual production and interactive dialogue. Recent advancements in Multimodal Large Language Models (MLLMs) have expanded the capacity of computing systems to interpret both textual and visual data, offering a new frontier for AI-mediated therapeutic support. This work-in-progress paper introduces an MLLM-powered chatbot that analyzes visual creation in real-time while engaging the creator in reflective conversations. We conducted an evaluation with five experts in art therapy and related fields, which demonstrated the chatbot's potential to facilitate therapeutic engagement, and highlighted several areas for future development, including entryways and risk management, bespoke alignment of user profile and therapeutic style, balancing conversational depth and width, and enriching visual interactivity. These themes provide a design roadmap for designing the future AI-mediated creative expression tools.
\end{abstract}


\begin{teaserfigure}
\centering
  \includegraphics[width=0.9\textwidth]{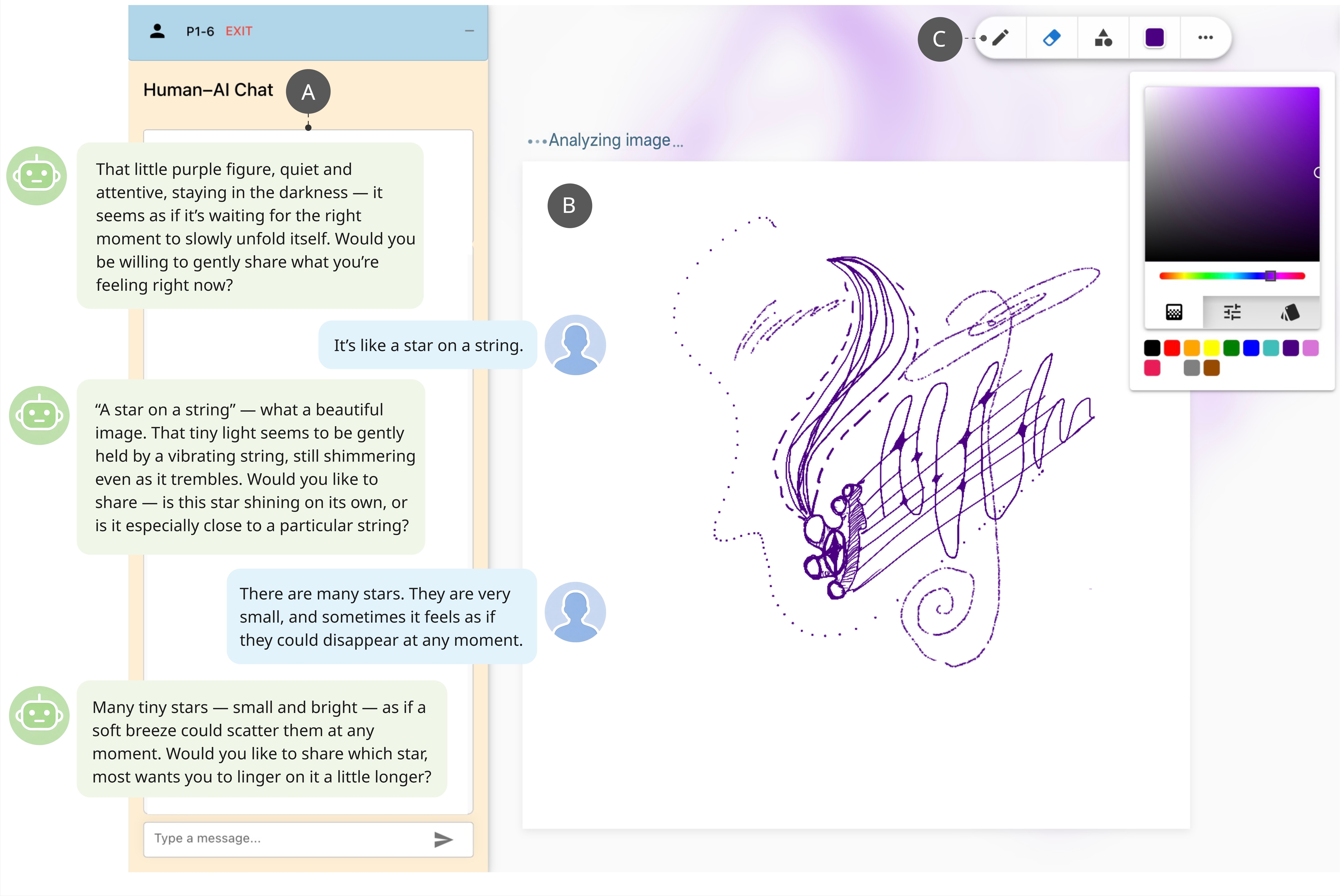}
  \caption{The overview of the AI-mediated therapeutic art activity platform: (A) Conversation Window where the user engages in emotional expression with a chatbot that  analyzes their drawing, (B) a Canvas that enables the user to create free drawings, and (C) a Toolkit that provides multiple visual creation tools. The screenshot is from our expert evaluation (P5), originally conducted in Chinese and translated into English for presentation.}
  \Description{Screenshot of the web-based AI-mediated therapeutic art platform interface. The interface is divided into three functional areas: (A) A vertical sidebar on the left displays the chat history between the user and the AI, including a text input field at the bottom. (B) A large central workspace features a white digital canvas displaying a purple line drawing, with a floating "Analyzing image" status indicator above it. In the top-right corner, a compact floating toolbar (C) provides visual creation instruments and a color palette for artistic expression.}
  \label{fig:teaser}
\end{teaserfigure}


\maketitle

\section{Introduction and Backgrounds}
Therapeutic art activities, such as expressive drawing and painting, are widely adopted for processing emotions and reducing stress, which have shown to improve the psychosocial health of individuals of all ages~\cite{feen1996art, weinfeld2024theoretical}. Unlike purely verbal expression, these activities leverage the visual metaphors for individuals to externalize complex internal thoughts that are often difficult to put into words~\cite{weinfeld2024theoretical}. 

The transition from simple mark-making to meaningful self-discovery typically requires the presence of a skilled practitioner, such as an art therapist~\cite{feen1996art, beaumont2012art}. These practitioners scaffold the experience by asking evocative questions, noticing patterns in the visual creations, and maintaining a safe space for the creator to share their thoughts~\cite{beaumont2012art}. Without this guidance, solo creation can feel aimless or fail to reach deep insights~\cite{santosa2026emoflow}.
While the importance of the human practitioner is undeniable, access to such specialized support remains a limited~\cite{uttley2015systematic}. 
Furthermore, social stigma often prevent many individuals from seeking human help~\cite{zheng2025customizing}.

Recently, the rapid development of language and vision AI models has advanced digital support from simple text‑based conversation to visual interaction\add{~\cite{zheng2025artmentor, li2025videoa11y, mingzhe2026adcanvas}}, opening up the opportunties for expanding the accessibility of therapeutic art activities and lowering the entry barrier associated with social stigma~\cite{lee2023exploring}.
For example, Yang et al. utilized a fine-tuned YOLO-v8 model to detect sketch content and generate feedback to support beginners of art therapists \cite{yang2025practicedapr}. 
Du et al. built a platform that supports users co-create digital painting with AI~\cite{du2024deepthink}. 
More relatedly, Liu et al. implemented TherAIssist, featuring a chatbot that guides art creation and encourages self-expression as part of therapist-customized homework. \add{However, these systems lack the capability to link the dialogue directly to the evolving visual metaphors on the canvas, preventing synchronized multimodal reflection that is essential in therapeutic art creation} ~\cite{liu2025theraissist}. 
\add{On the other hand, while some systems have incorporated multimodal information understanding, they are not designed to for therapeutic purposes~\cite{shi2026talksketch, huang2025sketchgpt, leng2025genfodrawing}. For example, Shi et al. focused on assisting designers in early-stage ideation through real-time speech and drawing inputs~\cite{shi2026talksketch}.}
\add{Peng et al. introduced a digital moodboard tool, DesignPrompt, which assists designers in formulating descriptive prompts for AI image generation by integrating diverse inputs such as colors, reference images, and semantic labels~\cite{peng2024designprompt}.}

\add{Although one recent system~\cite{wu2025navigating} combines a collage-like visual creation tool with visual understanding to generate customized AI responses for trauma-impacted youth, it restricts users to pre-defined assets restricts spontaneous artistic expression.} To address this gap, we developed a web-based \add{free-form drawing} system featuring a chatbot powered by Multimodal Large Language Models (MLLMs), which analyzes visual creations in real time and facilitates interactive conversations based on the visual information \add{in therapeutic art activity}.
As a preliminary step, we conducted qualitative evaluation with domain experts in art therapy or related fields (\textit{N} = 5): each participant was invited to interact with the system by simulating a typical therapeutic session, and then provide feedback on the practicalities and challenges of deploying such a tool in real world. \add{By evaluating our system with domain experts in the first step, we can mitigate potential risks associated with AI-driven therapeutic support before deploying the system to vulnerable target users~\cite{wu2025navigating}.}

Our \add{expert} participants recognized the chatbot's potential to facilitate therapeutic engagement, praising  its \add{real-time} visual comprehension and ability to manage a supportive conversation. However, the study also surfaced critical areas that must be addressed to ensure the safety and effectiveness of AI-mediated therapeutic tools, such as entryways and risk management, balancing the ``depth and width'' of inquiry, \add{tailoring the AI's therapeutic style to individual user profiles,} and opportunity to enrich the experience through direct, collaborative visual engagement between the user and the chatbot.

Our contributions to the HCI and DIS communities are two-folds: (1) a novel system pipeline that leverages MLLM to bridge the gap between creative production and verbal reflection \add{in therapeutic art activities}; and (2) insights from expert evaluation regarding the requirements for AI-mediated therapeutic art tools, which offer a roadmap for researchers and designers building responsible, context-aware agents.
\begin{figure*}[t]
\centering
\includegraphics[width=\linewidth]{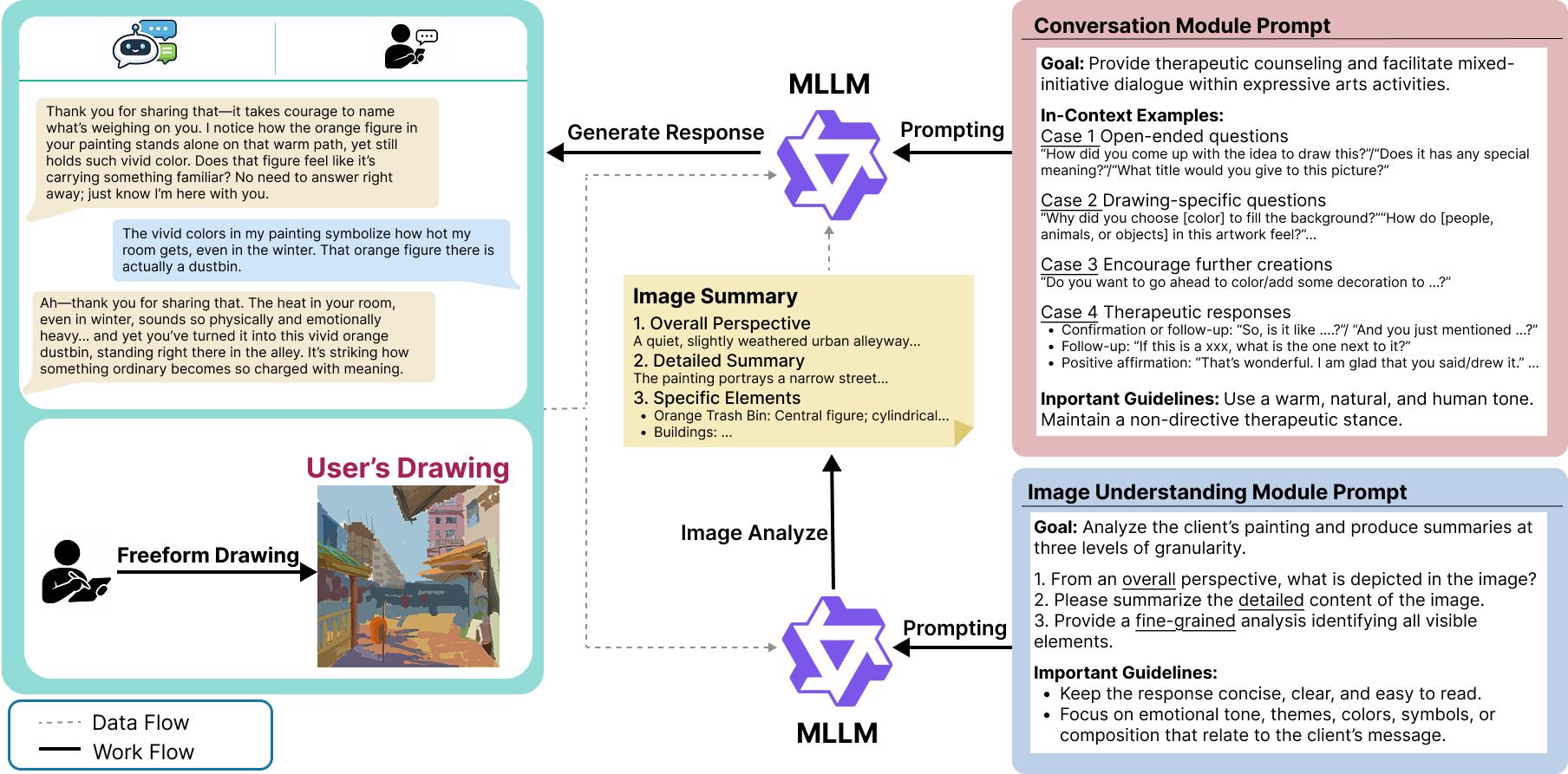}
\caption{An overview of the system architecture. 
The image understanding module takes the client's drawing and the designed prompts as input, producing an image summary that the conversation module uses to generate responses. }
\label{fig:system2}
\Description{An overview of the system architecture, mainly including a Conversation Module and an Image Understanding Module. On the right, two text boxes provide representative prompts used by each module. Users' drawing and conversation history are fed into both modules. The Image Understanding Module processes the drawing to generate a visual summary, which then serves as a contextual input for the Conversation Module to produce a coherent therapeutic response.}
\end{figure*}
\balance{}
\section{System Design and Implementation}
We developed a web-based system featuring a MLLM-powered chatbot to facilitate therapeutic art activity, which serves as a design probe for us to examine how AI analyze visual metaphors and sustain interactive dialogue. Below, we describe our design and implementation in detail.
\subsection{Design Rationale}


\subsubsection{Real-time drawing analysis without making assumptions.} 
To facilitate meaningful interactions, the system must provide responses grounded in the user's creative process~\cite{nash2020response}.
Thus, our system continuously monitors the strokes on the Canvas. To balance analysis precision and computational cost, the system calls the image-understanding module every three minutes, when the user saves the Canvas, or when it detects substantial pixel updates (cumulative color value changes across more than 20\% of the total canvas area). 
We explicitly instructed the image-understanding module to produce descriptive summaries (e.g., colors, positions) rather than speculative assumptions (e.g., what a specific visual implies) to avoid over-interpretation~\cite{dalley2008art}. 
\subsubsection{Promoting active expression while allowing silence.} 
The process of externalizing emotions through art is often accompanied \add{with} hesitation and uncertainty, especially for those who are new to art-making~\cite{santosa2026emoflow}. In these cases, a safe space should gently encourage the user to proceed, while allowing them to stay silent if needed~\cite{horvitz1999principles, zhang2026asafeplace}. 
Therefore, \add{in our system, while the chatbot responds immediately to any text inputs from users, it does not preemptively push users to progress their artwork if the canvas remains static. Only} if the user is stuck in the creation process without actively sharing their thoughts in the conversation over two minutes, the chatbot will provide ``gentle nudges'' that engage the user to share their thoughts or explore new visual elements. \add{If the user chooses to remain silence after the nudge, the system will suspend further autonomous messages, respecting the user's need for a safe, uninterrupted creation space.}


\subsubsection{Encouraging visual externalization.} 
Externalizing inner thoughts into visual metaphors is an important part of therapeutic art-making~\cite{briks2007art}. 
In a traditional session, a practitioner facilitates this by encouraging the creator to manifest their spoken reflections as physical changes on the canvas~\cite{briks2007art}. To emulate this process, we made the chatbot analyze the user's verbal narrative, and then prompts them to translate these insights into visual elements.
This rationale ensures that the dialogue is not merely an ``interview'' about the art, but an active driver of the creative process itself.

\subsection{System Overview}

\subsubsection{Front-end}
As shown in Figure \ref{fig:teaser}, upon logging in, the user is presented with a workspace consisting of an interactive drawing canvas and a conversational sidebar. They may initiate the session by drawing directly or by engaging the chatbot to brainstorm initial metaphors and concepts.

\subsubsection{Back-end}
To power the front-end interface, our back-end architecture incorporates two key modules, as shown in Fig. \ref{fig:system2}. Both modules are implemented based on the \textit{Qwen3-VL-Plus} model \cite{team2025qwen3}, leveraging in-context learning to optimize performance for therapeutic art tasks: (1) an \textbf{Image Understanding Module} that conducts real-time analysis of users' drawing content, for which we designed a structured prompt that instructs the model to generate three levels of interpretations of the user’s artwork, including an overall summary of composition and style, a detailed list of all drawing elements, and a fine-grained analysis of each element’s attributes (e.g., color, size, position); and (2) a  \textbf{Conversation Module} that integrates the conversational history, current drawing, and analysis results from the image understanding module as input, and synthesizes these contextual inputs to enable the chatbot to generate context-aware, therapeutically appropriate responses. To this end, we designed task-specific prompts and in-context examples tailored to therapeutic art contexts, which are engineered to encourage reflective expression, elicit open-ended and drawing-focused questions, and encourage further creation.  

\add{To ensure data security, all multimodal data (i.e., text and images) are stored securely on our private server. Each participant was assigned a unique username and password, with credentials encrypted in the database to prevent unauthorized access.}



\section{Method}
\setlength{\tabcolsep}{3pt}
{\sffamily
\begin{table*}[tbp]
\small
\centering
\caption{Participants' demographic information.}
\Description{A summary table of the five expert participants' demographics and professional backgrounds. The five columns display ID, Gender and Age, Background, Relevant Experience, and Specialized therapeutic areas.}
\label{participants_demographics}

\begin{minipage}{0.95\textwidth}
\centering

\begin{tabular}{l l p{4cm} l p{5cm}}
\toprule
\textbf{ID} &
\makecell[l]{\textbf{Gender/}\\\textbf{Age}} &
\textbf{Background} &
\makecell[l]{\textbf{Relevant}\\\textbf{Experience}} &
\textbf{Specialized Areas} \\
\midrule
P1 & F/31 & Healing Arts Practitioner & 1 year & Stress management \\
P2 & F/32 & Certificated Art Therapist & 1 year & Chronic mental condition management \\
P3 & M/23 & Healing Arts Practitioner & 1 year & Psychological projection; drawing- \& olfactory-based healing activity \\
P4 & F/27 & Healing Arts Practitioner & 3 years & Geopsychology; drawing- \& crafting-based healing activity \\
P5 & F/30 & \makecell[l]{Dancer \& graduate student in\\Expressive Arts Therapy} & 10 years/6 months & Emotion expression through dancing \& drawing activity \\
\bottomrule
\end{tabular}

\end{minipage}
\end{table*}
}

We conducted an expert evaluation with five participants: one certificated art therapist, three healing arts practitioners, and one graduate student in expressive art therapy.
Except for P3 who often used digital drawing tools in their practice (e.g., Tablet, VR), all other participants primarily practiced therapeutic art in physical settings.
As shown in Table \ref{participants_demographics}, participants' ages ranged from 23 to 32 (\textit{M} = 28.6, \textit{SD} = 3.26). 
The study occurred online or offline depending on each participant's preference, and lasted about two hours. Offline participants were provided with an iPad and an Apple Pencil for a smooth drawing experience; online participants used their own computer or iPad. 
The study was conducted with each participant individually in three stages as described below:

\begin{itemize}
    \item \textbf{Introduction} (15 mins): We began by introducing the study's goals and procedures to the participants, and also asked about their training backgrounds and common workflow.
    \item \textbf{Prototype Exploration} (15-20 mins): Following a brief system demonstration, each participant was asked to freely explore the system as a client in an therapeutic art activity. \add{We encouraged free exploration to capture how participants naturally interact with the prototype, while gently prompting them to talk with the chatbot if they merely focused on drawing to ensure all the features were tested~\cite{rubin2008handbook, blandford2016qualitative}. Participants} were invited to think-aloud during the interaction, sharing their perspectives of the chatbot's response quality, dialogue pacing, visual comprehension, and any communicative friction or misunderstandings.
    \item \textbf{Debriefing Interview} (40-60 mins): Reflecting on the prototype exploration, we asked participants about the feasibility of integrating an AI-mediated system into their professional practice. Specific attentions were paid to additional features they hope to add and ways to handle potential risks.
\end{itemize}

\add{\textbf{Ethics Statement}: This study was approved by our institutional ethics review Board, with informed consent obtained from all participants. During the study, we emphasized that our prototype was an assistive therapeutic tool aimed at fostering self-expression and reflection rather than a formal clinical instrument or a replacement of professional clinical care.}


\section{Findings}

\begin{figure*}[t]
\centering
\includegraphics[width=1\linewidth]{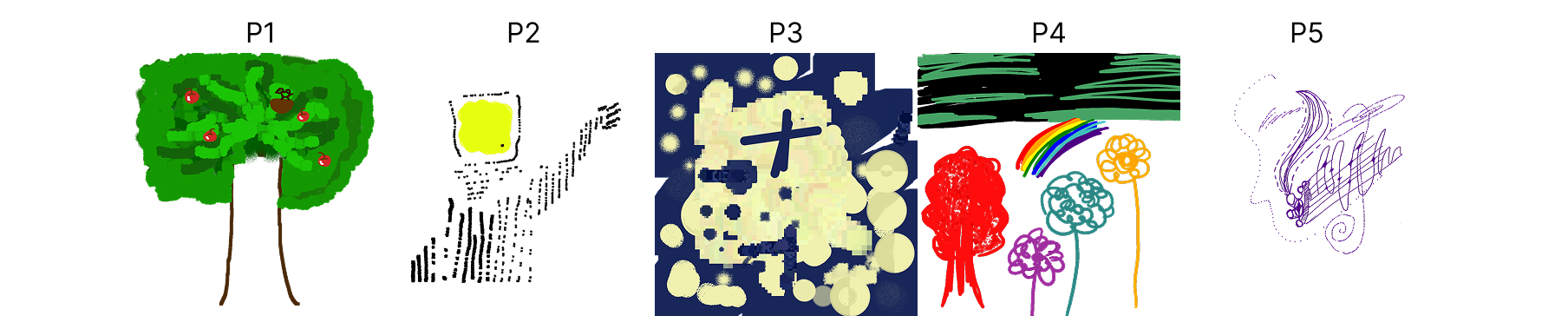}
\caption{The artworks created by the expert participants using our system during the study.}
\label{fig:paintings}
\Description{A collage of five diverse artworks created by expert participants. P1 created a green tree with red apples. P2 created a humanoid dotted pattern with a square, fluorescent yellow face. P3 painted a deep blue background centered with a dense area of textured yellow-green dots and a modified deep blue cross. P4 depicted a black top section with green lines and a white bottom section featuring a row of multicolored flowers and a small rainbow. P5 used abstract purple lines on a white canvas to draw a stylized musical score and notes, joined by four-pointed stars at the intersections.}
\end{figure*}


In general, our participants shared a positive outlook regarding the role of MLLM AI in therapeutic art activities: \q{AI can bring art to everyone's life, fostering their perceptual and aesthetic awareness for wellbeing} (P1).
They recognized the chatbot's potential to serve as a \q{low-stakes entry point} for emotional expression through art-making (P1, P3--P5), praising its visual comprehension, and the ability to stay open, neutral, and encourage self-expression. 
\add{For instance, P3 noted that the chatbot's \q{precise analysis of the visual creation}, which \q{deepened the conversation in a meaningful way}. }
\add{P4 found that questions asked by the chatbot were grounded in the visual details, such as color and shape of each elements, making the interaction feel personalized. }
\add{Similarly, P5 valued the system's acute sensitivity to the ongoing creation process. noting its ability to validate even accidental interactions. When the participant inadvertently placed a small purple dot on the canvas, the chatbot immediately integrated it into the dialogue rather than ignoring it or treating it as an error: \q{I might have touched the canvas by mistake and left a tiny purple dot... but the AI said, 'I see you've drawn this little purple dot, letting it stay in the dark.' I think this is excellent for someone who feels they 'cannot draw.' It doesn't judge; it doesn't say you haven't drawn anything. It finds value in even the smallest mark}.}
\add{Interestingly, P4 envisioned that the real-time image analysis capability could serve as a training aid for novice art therapists. }


On the other hand, they all highlighted the importance for mitigating potential risks such as inadvertently triggering severe emotional disturbance. While AI can understand the image content, currently it cannot truly ``observe'' the creation process---such as the creator's facial expression, body movement, etc (P1, P3). Furthermore, the conversation lacked the depth required for self-discovery (P2), and the digital experience often misses the tactile and sensory richness in physical creation (P1, P5). In light of these challenges, they pointed out several areas for future improvement.

\subsection{Entryways \& Risk Management}
All participants suggested a screening process to exclude individuals, for whom unguided deep-diving into visual metaphors could lead to emotional dysregulation: \q{there should be a pre-assessment, like a short survey, to help determine whether the client is suitable for using AI} (P4). 
When discussing potential risks, P4 suggested that the system should monitor for signs of therapeutic rupture---instances where the user becomes frustrated, angry, or feels misunderstood by the AI:
\q{If the user had rejected the AI’s suggestions, or if the system can detect signs of anger or dissatisfaction in their emotions, it may be the right moment to introduce human interventions.}

\subsection{Bespoke Alignment Between User Profile and AI's Therapeutic Style}
Participants emphasized that a ``one-size-fits-all'' AI persona cannot meet the diverse user needs and proposed two dimensions of personalization. First, art therapy is grounded in a wide range of theoretical frameworks, and the user should have an option to choose which style they prefer:
\q{Some therapists follow psychodynamic approaches, some are humanistic, others are narrative, and some practice cognitive behavioral therapy (CBT). Users may care about the therapist’s orientation when making their choice} (P4). 
P5 shared a related example from the user’s perspective: \q{If I were a user and a dancer, I would prefer to choose a therapist who understands dancing. Otherwise, they might not fully grasp my experiences.}

Second, P2 mentioned a user profile of their basic demographics and communication style can help her get prepared before the sessions and avoid inappropriate language. P1 and P4 further shared the importance for AI to have a long-term memory to establish ``\textit{alliance}'' with the user. Otherwise, the interaction can feel transactional rather than therapeutic. 
\subsection{Balancing Therapeutic Depth and Width}
All participants hope\add{d} the chatbot could ask more details about the visual elements they created on the Canvas. Particularly, P1 and P5  mentioned that the questions should move from the big picture to smaller pieces, which can help lower the user’s defenses, enabling a continuous exploration of their thoughts.
Beyond deep reflection, P2 highlighted the necessity of ``\textit{conversational width}.'' This could give users more space for imagination rather than limiting their thoughts in one box, and for people experience acute mental health distress, this can \q{avoid overstepping into potentially triggering territory} (P2).
\subsection{Enriching Visual Interactivity}
During the debriefing interviews, participants expressed interest in enhancing the art-making process through contextual UI interactions and human–AI co-creation. For example, P5 suggested that the AI chatbot could engage with specific visual elements by highlighting or masking regions directly on the canvas for discussion, rather than putting them in pure language description, as visual communication is an important part of drawing-based art therapy. 
P2 added, \q{Compared to text, I personally tend to respond more naturally to visual elements.} 

\section{Future Work}
By combining real-time visual analysis with a supportive conversation, our system demonstrate the potential of AI to function as a low-barrier entry point for therapeutic art activity. 
At the same time, our expert evaluation warned several areas of improvement. Going forward, we will focus on refining the chatbot's conversation flow to balance the depth and width of the interaction, integrating visual interaction features that allow the AI to move beyond text and directly communicate with the user on canvas. We will develop a risk management module that monitors for signs of distress, ensuring a safe and responsible wellbeing support. \add{Prior to real-world user studies, these additions will need to be validated by more expert therapists with diverse training backgrounds and practicing experiences.}

\add{Looking ahead, our research will transition from this initial expert validation to a multi-phased user studies. We will begin with controlled lab studies to ensure the safety and usability of the system, and then proceed with longitudinal field studies to observe how this the system integrates into people' daily lives.
By gathering granular interaction patterns over time, we aim to understand how these multimodal interactions facilitated by AI shape users' day-to-day emotional expression and self-reflection processes. Throughout the study, we will employ a proactive monitoring framework to identify and mitigate potentially triggering content, ensuring participant safety remains paramount. To provide a comprehensive evaluation, we will triangulate this qualitative interaction data with quantitative measures~\cite{betts2006art, saunders2000evaluating}, allowing us to measure the system's long-term efficacy in fostering psychological resilience. }


\begin{acks}
\add{We thank our participants for their time and interests. This project was supported by the Innovation and Technology Fund (ITF) of the Innovation and Technology Commission (ITC) of the Hong Kong Special Administrative Region (HKSAR) Government (\# ITS/269/24FP) and City University of Hong Kong (\# 7020106). }
\end{acks}

\bibliographystyle{ACM-Reference-Format}
\bibliography{sections_revised/reference} 



\end{document}